# PRT (PERSONAL RAPID TRANSIT) NETWORK SIMULATION

*Włodzimierz Choromański, Warsaw University of Technology, Faculty of Transport, prof.wch@gmail.com*

*Wiktor B. Daszczuk, Warsaw University of Technology, Institute of Computer Science, wbd@ii.pw.edu.pl*

*Jarosław Dyduch, Warsaw University of Technology, Faculty of Transport, slavypl@gmail.com*

*Mariusz Maciejewski, Warsaw University of Technology, Faculty of Transport, Mariusz.Maciejewski@kontron.com.pl*

*Paweł Brach, Warsaw University of Technology, Faculty of Transport, braszek@gmail.com*

*Waldemar Grabski, Warsaw University of Technology, Institute of Computer Science, wgr@ii.pw.edu.pl*

## ABSTRACT

Transportation problems of large urban conurbations inspire search for new transportation systems, that meet high environmental standards, are relatively cheap and user friendly. The latter element also includes the needs of disabled and elderly people. This article concerns a new transportation system PRT - Personal Rapid Transit. In this article the attention is focused on the analysis of the efficiency of the PRT transport network. The simulator of vehicle movement in PRT network as well as algorithms for traffic management and control will be presented. The proposal of its physical implementation will be also included.

## INTRODUCTION

In Warsaw University of Technology, a project on Personal Rapid Transit (PRT) (Irving at al. 1978, Andreasson 2010) is under development (Choromański et al. 2011a and 2011b, Daszczuk at al. 2011).
Personal rapid transit (PRT), is a public transportation mode featuring small automated vehicles operating on a network of specially-built guide ways. PRT is a type of automated guideway transit (AGT), a class of system which also includes larger vehicles all the way to small driverless subway systems. The whole system is electrically supplied.

In PRT designs, vehicles are sized for individual or small group travel, typically carrying no more than 3 to 4 passengers per vehicle. Guide ways are arranged in a network topology, with all stations located on sidings, and with frequent merge/diverge points. This approach allows for nonstop, point-to-point travel, bypassing all intermediate stations. The point-to-point service has been compared to a taxi.

There are three kinds of nodes in the PRT network: stations, capacitors and intersections.

Stations are places where passengers book their trips and board the vehicles, or wait for vehicles in a queue if there are no empty vehicles on a station. A capacitor is a source of vehicles (and sometimes may serve as a parking place). Intersections are threefold: "fork" (diverge), "join" (merge) and "junction" (such intersections are for technical purposes only).
It is assumed that a vehicle has its own control unit, which is linked via radio network with control units of other vehicles and nodes (capacitors, stations and "join" intersections). Radio connections are established to vehicles and nodes that are closest to the vehicle, i.e. not farer that specified distance. This makes a subnet "visible" to the vehicle, the edge of which is called "a horizon". The horizon distance should be chosen carefully, too large causes too much information to be transmitted (and routing problems), too small results in reduced safety of the traffic.

A vehicle gets information about current parameters of movement of preceding vehicles: their positions, velocities and mode of operation (acceleration/ constant velocity/ deceleration/ friction braking). From an intersection controller a vehicle receives the decision on priority of crossing the intersection.
Among other mechanical, electrical and transportation research goals, simulation of PRT network is performed, on two abstraction levels:
- Coordination level is "behavioural simulation". On this level, algorithms for following a route, keeping up, coordination on "join" intersections, joining the traffic and similar are tested for effectiveness.
- Management level is "statistical simulation". Simulation experiments identify the impact of various parameters of management algorithms (mainly of empty vehicle management and dynamic routing) on the passenger comfort (trip time and queue size).

There are several PRT simulators available (Castangia and Guala 2011, Zheng, Jeffery and McDonald 2009, Andreasson 2010, Hermes, Beamways, RUF), yet the authors have decided to build the project's own simulators, for two reasons:
- Available simulators have fixed, inaccessible traffic algorithms (on both levels: coordination and management); the user can only observe the operation of included algorithms and he/she has no influence on their structure and parameters.
- Simulators produce a number of synthetic output parameters; there is no access to individual events (witch should be reported as a kind of "event log"); this lack does not allow the researcher to build his/her own detailed characteristics of the traffic.

It seems that the own simulators build are more advanced. They permit the analysis of various algorithms of coordination and management, and they give access to wide and detailed set of traffic parameters and event logs. The purpose of simulators is:

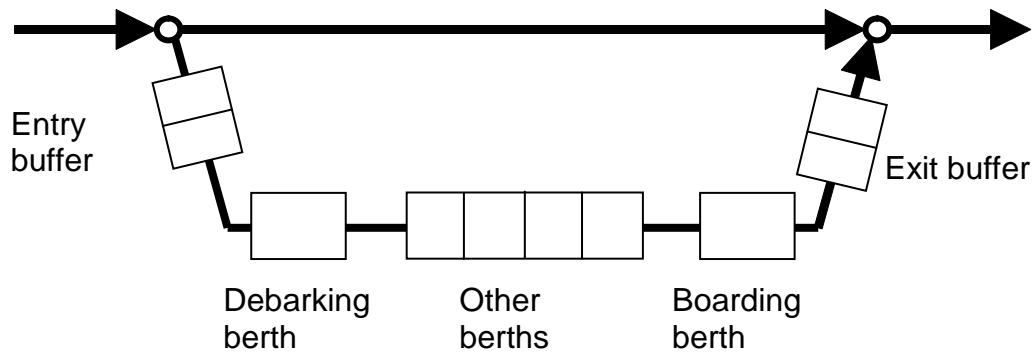

**a) in-line**

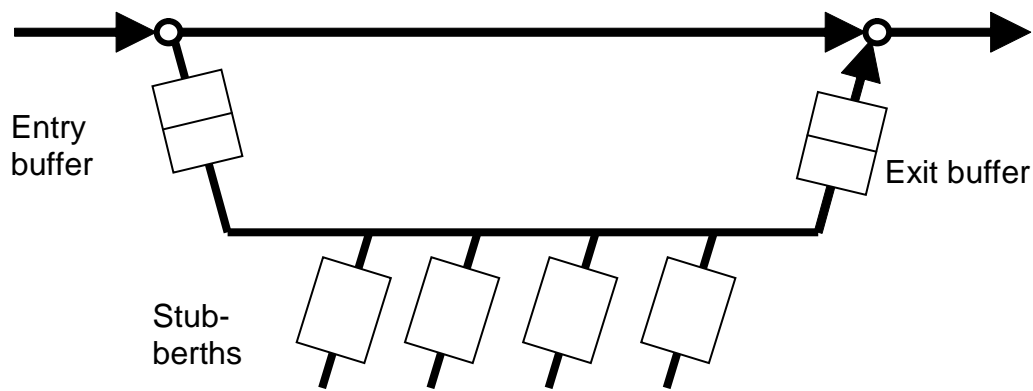

**b) stub-berths**

Figure 1 – station types

- analysis of various network topologies of PRT,
- identification of capacity and ridership of various networks,
- identification of saturation points (when performance is degraded to inaccessible level)
- research on dynamic algorithms (especially empty vehicles management and dynamic routing), and identification the operational cost of performance improvement (number of empty trips, distance of empty movement etc)
- sensitivity of the system to variation of different traffic parameters values.

## THE STRUCTURE OF THE PRT NETWORK

A structure of PRT network is a graph (Andreasson 2010). Nodes are capacitors, stations and intersections. Edges are segments.

There are two types of stations: in-line (with the FIFO rule applied) and with stub-berths. Vehicles must have ability to drive backwards to operate inside a berth. A structure of both types of a station is shown in Figure 1.

From topography point of view, nodes are simply points of size zero. Yet the nodes have internal structure necessary to model their roles in a network. The stations are most complicated: they have parking places (berths), entry and exit buffers, passenger queues etc. Several parameters characterize a station, including its type (in-line/stub-berths), number of berths, sizes of entry and exit buffer. A capacitor is much simpler, because it is not related to passengers. It has only one characteristic – a number of parking places.

Capacitor, station and "junction" intersection have one entry and one exit. "Fork" intersection has two exits (to diverge) while "join" intersection has two entries (to merge). Segments are edges of PRT graph. Segments are unidirectional and they have starting and ending points in nodes (capacitors, stations and intersections). A segment is characterized by its length and maximum velocity.

Nodes and edges conform a static structure of PRT network. Two types of dynamic objects define a behaviour of the network: vehicles, which originate from capacitors, and passengers appearing in stations. A Passenger group has a specific destination station. Passengers wait for the vehicle (or take an empty one, standing in berth), then a trip is realized, and after completing the journey, they ''disappear" from the system.

The basic goal of PRT network is to realize trips between stations for passengers. A vehicle is characterized by: capacity, maximum velocity, maximum acceleration and deceleration, maximum friction deceleration (emergency brake) and minimum (static) separation between vehicles. Passengers are organized in groups for a trip to common destination (station). A group is characterized by its cardinality and the target station.

## THE MODEL

The control algorithm of the PRT network is specified in two levels: lower - coordination level, and upper - management level:
- Two coordination algorithms are defined: keeping up, and coordination on "join" intersections (joining the traffic from a capacitor/station is realized as a special case of coordination on "join" intersection).
- Several management algorithms, including some empty vehicle management algorithms and dynamic routing algorithm.

The coordination algorithms use traffic parameters like maximum velocities, maximum acceleration and deceleration, separation between vehicles etc.

Some basic rules of behaviour are fixed in the simulator, including movement inside stations, and movement along the track and keeping up rules, are fixed in the simulator. The routing is based on Dijkstra's algorithm, although edge weights in the algorithm are parameterized by several values depending on static structure (segment lengths), traffic parameters (maximum velocities) and dynamic traffic situation (traffic jams).

Various characteristics of the model have different variability:
- The structure and topography of the network, type of individual stations (inline/stub berths), capacity of nodes, number of vehicles, boarding and debarking times as well as distribution of passenger group cardinality are constant.
- Traffic parameters stay constant during single simulation.
- Mean input of passenger groups and origin-destination matrix may be defined to be valid in specific periods.
- Maximal velocity of a segment may be changed in specified while.

# TWO APPROACHES TOWARDS PRT SIMULATION

Two independent simulators have been built: one event-driven and second based on cellular automata. Both simulators use the same model of PRT network and the same base parameters. This allows to compare the results and verify correctness of the two simulators. The simulators have different usability properties: the former simulator is more accurate, while the latter one is much faster.

Both simulators use micro-simulation (Daszczuk at al. 2011, Fox 1999, Barcelo 1999, Gabard 1999, Nagel and Schreckenberg 1992). In the event-driven simulator, segments are divided into sectors and passing a sector is a basic simulation step. The second simulator is based on cellular automata, where every automaton corresponds to a singular node of the network. The basic step of simulation is 1 second interval, in which all automata are updated.

The design of the model typically follows a hierarchy of design steps, from most general to most detailed:

- network structure and topography;
- characteristics of network elements: station types, numbers of berths, maximal velocities of segments etc.;
- acceleration and deceleration, general velocity limits, separation, boarding and debarking times, origin-destination matrix;
- parameters of empty vehicle management, priority rules and dynamic routing;
- number of vehicles and their individual characteristics (velocity limits for individual vehicles);
- passenger input distribution.

The simulation experiments may be used to compare throughput characteristics of various network structures, various station and segment parameters, or particular network in which empty vehicles management parameters or dynamic routing parameters vary. Ability to perform the same experiments on two different simulators assures results verification that will be model independent. Some research is described later.

# EVENT-DRIVEN SIMULATION ENVIRONMENT

The movement of vehicles is performed in the event-driven manner, which is a typical solution (Anderson 1998, Lopez et al. 2008). Every segment is divided into sectors. The decisions on the behaviour of vehicles are made in the connections of sectors. Also, starting some actions in capacitors/stations – passenger group occurring, coupling of passenger group with a vehicle, beginning of boarding or debarking, start form parking place etc. are simulation events.

The simulation environment is called Feniks 3.0. The main program of the simulator (called "UI") is responsible for building a model, setting its parameters and to organizing simulation experiments. The external library (called "external simulator") implements control algorithms (on both layers). The whole simulator is written in C#. The interface is defined between the simulator parts, which allows to define user's own control algorithms instead of supplied with external simulator.

Every event (for example start of movement, time of passing the sector) is served by calling external simulator and performing adequate acts regarding obtained decision. The main program (UI) is a simulation engine, an animation engine and a statistics collector. Also, in UI are included procedures for defining a model structure and topography. External simulator decides on every change of state of every vehicle, for instance:

- coupling of passenger group with a vehicle;
- beginning of boarding or debarking, passing a connection of sectors, stopping at connection of sectors;
- decision on the time of passing the sector; the time is calculated by the coordination algorithm that guarantees safe distance between vehicles on a track.

The Simulator plans the vehicle movement analytically and therefore precisely (not approximate). Although the simulation is event-driven, the result of simulation is similar to the analogue simulator. The only inaccuracy comes from the fact that a vehicle may stop only at the connection of sectors (rather than inside a sector).

## Coordination - movement along a track

Vehicles move along a track with maximal possible velocity (which is a minimum from maximum velocities of a model, a segment, a sector and a vehicle). A "horizon" is defined to be a distance "observed" by the vehicle. The nodes and other vehicles closer than the horizon influence the behaviour of the vehicle.

The velocity on an empty track (in a distance closer than the horizon) is limited by the acceleration of the vehicle and by deceleration of the vehicle together with velocity limit on the next sectors. If there is a preceding vehicle on a track, or a node (station/capacitor/"join" intersection) which cannot be crossed because of traffic condition, a distance called "separation" is kept. Static separation defines the minimum distance. The coordination algorithm guarantees stopping after the preceding vehicle in static separation distance if preceding vehicles starts to decelerate of brake at the moment.

If there is a "fork" intersection closer than the horizon, only the chosen outgoing track (left or right) is analyzed.

If a node (capacitor or station) is closer than the horizon, and the node is the target of the trip, the vehicle must reach the zero velocity in a point of diverging from the track towards the node. The exception is when all the parking places and all entry buffers in the node are occupied – in such a case the vehicle must stop in separation distance before the node.

If a vehicles wants to "drive through" a node, it runs with maximal allowed velocity, provided that no vehicle starting from the node is in conflict (this case will be described in the next subsection).

## Coordination - behaviour on "join" intersections

A "join" intersection is allocated to one of approaching vehicles by the intersection controller. Only the vehicles closer then the horizon distance are considered. And only closest vehicles on both segments are taken into account, vehicles following them follow normal keeping up. From a vehicle's point of view, if there is a possible conflict:

- if the intersection is allocated to a vehicle on the other track leading to the intersection, the considered vehicle is planned to stop prior to the intersection in static separation distance;
- otherwise the vehicle ask the intersection controller which vehicle has the priority; if the other one – the considered vehicle stops as above.

Joining the traffic (from a capacitor or a station) is very similar, just the point of the capacitor on the track (or the point of the station) is treated as intersection, and it is being allocated to a vehicle on the move or to a vehicle joining the traffic, depending on priority rules (parameters of coordination algorithm).

## CELLULAR AUTOMATA SIMULATION ENVIRONMENT

Cellular automata is a structure defined by a matrix of cells and their states, transitions and the rules of those transitions (Nagel and Schreckenberg 1992). Automata in such form are mathematical models that construct an environment for a bigger, discrete classes of models, because all the structures describing them are discrete.

Each simple cellular automaton consists of n-dimensional, discrete matrix of cells. Each cell is the same (is a copy of the previous cell) and the whole space of the matrix must be filled with the cells put next to each other. Each cell has exactly one state from the finite number of available states. Transition of each cell takes place based on the same, precisely defined local rules (homogeneity), that depend only on the previous state of the cell and states of finite number of neighbouring cells. Transition is discrete and happens at the same time for all cells (parallelism). In the cellular automata, a cell is finite automaton.

In order to model the PRT network and traffic a more elaborated adaptation of cellular automata has been chosen – directed graph that represents the infrastructure.

The computational model is a directed graph, in which the nodes are the hubs and the edges between the nodes are the segments. Each node and edge has all the parameters that describe a given element (length of the segment, direction of movement, maximum allowed speed, etc.). Each edge is tied to a discrete model of a segment, which is represented by 1-dimensional array. One cell represents one unit of segment and is a parameter of the model.

Each junction is represented as one cell. In a given unit of time, in a given cell, there can be only one vehicle. The cell can have one of two available states – empty or occupied by a vehicle. Each vehicle in the model moves with a velocity from the range 0…V-max.

In the simulator a topographic model has been implemented, that consists of 2-dimensional, regular and discrete cell matrix. This model is a layer of abstraction over the directed graph. In the graph the nodes are the elements of the segments and the edges define the direction of movement between the nodes. Each node represents exactly one cell in the 2-dimensional matrix.

The configuration describing the infrastructure and the initial state of the environment – location of the vehicles, stations and capacitors and the passengers is one of the parameters of the model. The simulator uses and optimal-path algorithm to define a route to the target station. Such an approach allows for a dynamic control of the vehicles during the movement and is an excellent template of the real life movement of the PRT vehicles.

**Movement in the network**

After defining all the elements of the cellular automata, the rules of transition can be applied on the cell matrix. The transition process can be split into several parts (Schadschneidert and Schreckenbergt 1993). Initial state, as mentioned previously, is a the definition of the initial conditions of the cell. Usually those are neutral states, that do not cause any conflicts.
Update of the automata matrix is a walkthrough sequence of steps for each cell according to the instructions below:

1. Verification of the transition rules – in this step a current state of the cell is verified along with the states of the neighbouring cells and other parameters of automaton

2. Neighbours verification – during this step a verification of any conflict states of the neighbouring cell takes place. If there are any conflicts they have to be resolved according the predefined rules

3. Verification of boundary conditions – verification of the cell at the borders of the matrix. They can be removed (absorbing closed neighbourhood) or new ones can be created (periodic neighbourhood)

4. Verification of number of iterations- if this is a finite automaton, with predefined lifecycle than in this step it is check whether the transition should stop.

In the case of the implemented PRT network prototype, updated model of the classic cellular automaton has been implemented (Benjaminy et al. 1996, Li and Wu 2001). The model consists of the following steps (each is performed in parallel for all vehicles in the environment):

1. Acceleration: if the speed of the vehicle *V* is smaller than the maximum, allowed speed (for a vehicle or road segment) and if the distance to the next vehicle is bigger than *V+1*, than the velocity is increased by 1, e.g. *V:= V + 1*

2. Deceleration: if the vehicle at the position *I*, with velocity *V*, sees a vehicle at the position *i+j*, for *j* smaller or equal to v, then the speed is reduced to *j-1*, e.g. *V:= j-1*

3. Randomization (optional): with a probability $p_1$, the velocity of a vehicle is reduced by *1* (if bigger than *0*), e.g. *V:=V-1*

4. Randomization (optional): with a probability $p_2$, vehicle breaks down for a time period *J*, e.g. the velocity of vehicle is *0* for *J* units of time

5. If in the next unit of time vehicle will drive through a junction, following conditions are verified:

    a. If there is no conflict on the junction, e.g. there is no other vehicle coming to junction in the same time do nothing.

b. Otherwise define the order of the vehicles (using weights-based function). The priority vehicle does nothing (drives through the junction) and the other one slows down, letting the priority vehicle to drive through. The order of passing vehicles on the junction is defined by the weight of each vehicle - vehicle with higher weight goes first on the crossroad. The weight of the vehicle is determined based on the following function:

$$W(p) = W_t \times t + W_d \times d + W_p \times p + W_{pas} \times pas \qquad (1)$$

where:
*W(p)* – weight of the vehicle
$W_t$ - weight of waiting time
*t* – waiting time
$W_d$ - weight of the priority of the segment
*d* – priority of the segment
$W_p$ - weight of the priority of the vehicle
*p* – priority of the vehicle
$W_{pas}$ - weight of the number of passengers
*pas* – number of passengers in the vehicle

6. Movement: move vehicles *V* cells in the direction of movement

## SIMULATION RESEARCH

As an example of research work, a simulation experiment with empty trips in PRT network is presented. A PRT network performs people transport on their demand. Yet, to achieve the main goal, some trips without passengers (called *empty trips*) must be executed. The examples are:
- When a passenger group (typically 1-4 persons) occurs at a station, and there are no empty vehicles available in the station, an empty trip must be organized to supply a vehicle for the passenger group. It is the *calling* mechanism.
- When a vehicle approaches a station, and there is no empty berth for it, one of empty vehicles standing in the station must be *expelled* from the station.
- Whether there are many empty vehicles in a given station, and there are no (or little) empty vehicles in other station, an empty trip may be organized from the station to the latter one (target station), especially if average passenger input at the target station posits to expect new group soon. The mechanism is called *balancing*.
- If an empty vehicle stays in the station for a long time, and there are no passengers waiting or expected at the station, the vehicle may be *withdrawn* to a capacitor for safety reasons (especially at night).

The former two mechanisms (calling and expelling) are necessary in PRT system, while the latter two (balancing and withdrawing) are optional.

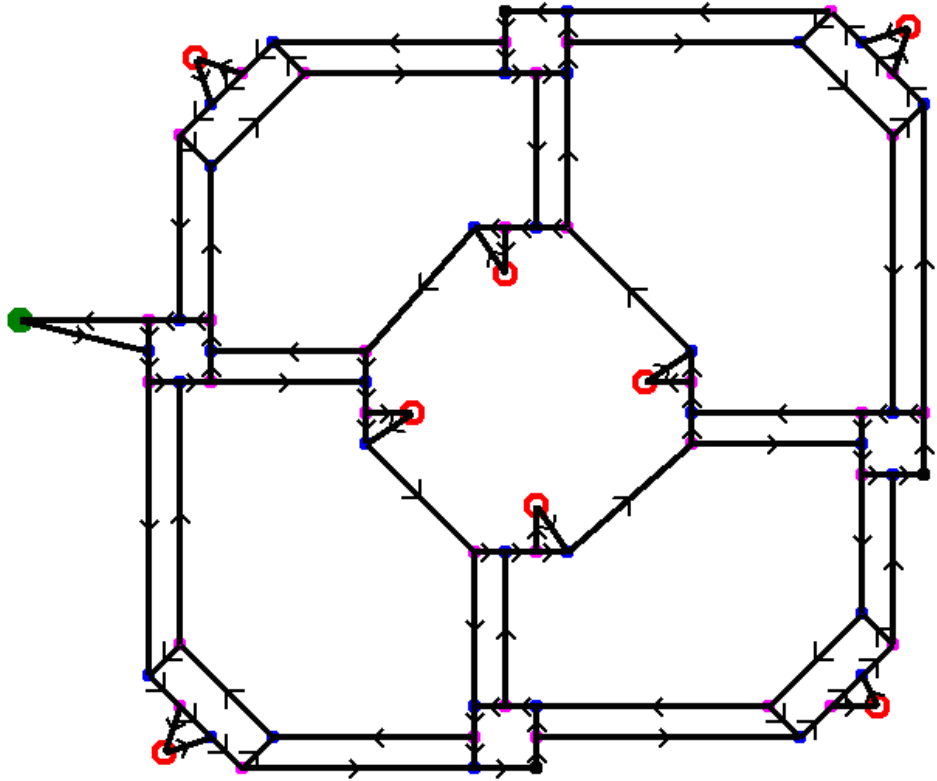

Figure 2 – The first PRT model (*city*)

In a simulation experiment, an impact of a passenger input rate on a number of empty trips is examined. As a PRT network is complicated and the influence of many design parameters in its behaviour is unknown, the research results may be obtained by simulation only. To show that a found relation is general, various network structures and various behaviour parameters should be applied.

For the research on the problem mentioned above, two PRT network structures were used, presented in Figures. 2 (*city*) and 3 (*seashore*). The circles in the figures represent stations while solid circles represent capacitors. Two numbers of vehicles were used. The main parameters of the network are:
- total tracks length 6064,5m (first model) or 5584m (second model),
- 12 or 24 vehicles,
- 4 passengers in every trip,
- all stations of in-line type,
- 4 berths in each station.
- maximal velocity 14m/s,
- maximal acceleration and deceleration 2m/s$^2$,
- boarding and debarking times 10s.
- static separation 4m (event-driven simulator) and 2m (cellular automata simulator),
- passenger input rate differs in specific simulations,
- trip destination chosen randomly

- empty vehicle management algorithm: calling, expelling and withdrawing mechanism used;

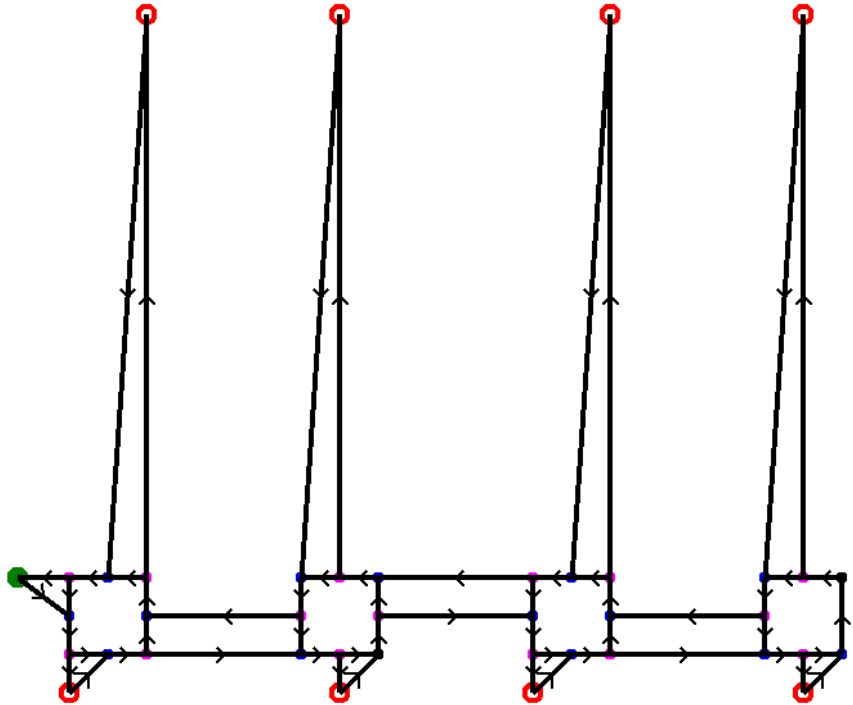

Figure 3 - The second PRT model (*seashore*)

- calling principle: nearest empty vehicle is called, if any;
- expelling principle: expelled empty vehicle is moved to a station on which there are empty berths and w distance to which is close (weighted expelling rule);
- withdrawing principle: if a vehicle stays in a station for longer then 120s, it is moved to the capacitor (timeout 120s), provided that there is an empty berth in the capacitor. During this time it may be called to other station (in which a passenger waits) or expelled (if other vehicle approaches the station and there is no empty berth for it).

The models (as with every model) have their saturation points, i.e. maximum numbers of trips per hour (depending on the model structure, number of vehicles etc.). The saturation point defines maximal throughput of the network: maximal number of passengers that may be moved to their destinations. For the research purposes, passenger groups input values were applied at saturation point and at several (5) input values less than maximum (passenger group input greater than in saturation point is useless because passenger queues rise to infinity).

The results obtained from event-driven simulator are collected in Figure. 4, for the two PRT structures and for two numbers of vehicles. The results from cellular automata simulator are collected in Figure 5. Experiments in both simulators give similar results, where in every plot there is a maximum of empty trips (though the input rates of the maximum and the maximum values differ).

The explanation of the observed feature is simple, although it is hard to expect it without simulation experiment. The reason of the plot shape is as follows:

- for input rates near saturation point every vehicle freed by passengers is taken immediately by successive passenger group, therefore empty trips seldom are obeyed;
- for low passenger input rates vehicles are seldom called (at most as many times as new passenger group occur at the stations); situations requiring expelling occur rarely (because of low number of trips with passengers); withdrawing may occur but at most as frequent as trips with passengers;
- if passenger input grows then the numbers of calling situations and expelling situations grow as well, while the number of withdrawing situations falls; the growth of empty trips is limited by coincidences of passenger occurring with empty vehicle standing in the same station; the coincidences lastly dominate and the total number of empty trips begins to fall.

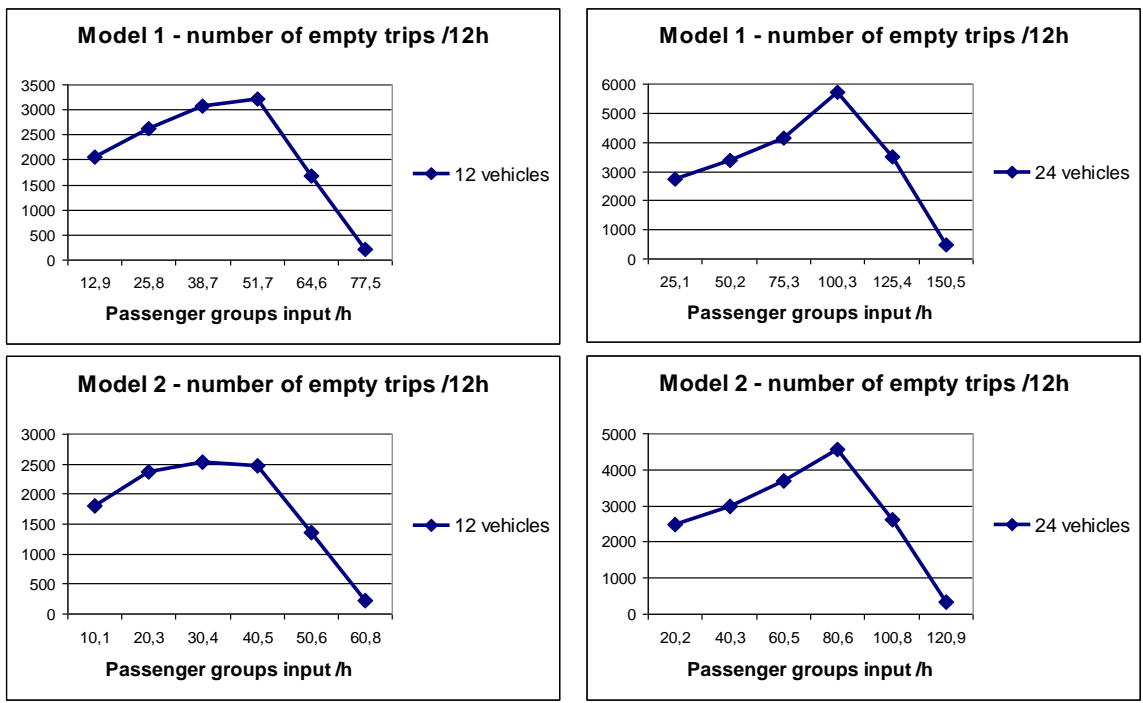

Figure 4 – Simulation results for both models and two numbers of vehicles (event-driven simulator)

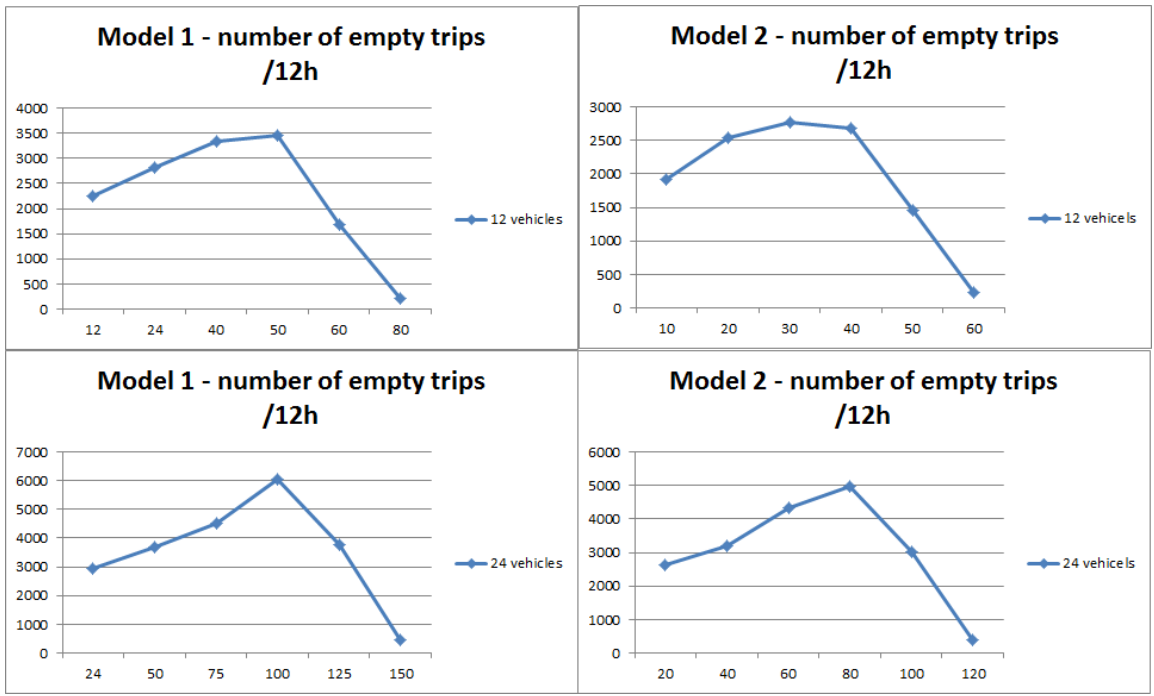

Figure 5– Simulation results for both models and two numbers of vehicles (cellular automata simulator)

It is obvious that the described relation takes place in every reasonable model (perhaps excluding pathological ones, for example if calling a vehicle to other station precedes taking empty vehicle standing in the same station as passengers occur).

## IMPLEMENTING PRT STEERING AND MANAGING ALGORITHMS IN PHYSICAL LABORATORY ENVIRONMENT

In parallel with developing logical model of the PRT network, a physical one is being built on the Warsaw University of Technology. It implements the algorithms described in the former chapters. The security and safety of PRT system is governed by a central computer control and dispatch system. The system consists of layers containing following subsystems (Figure. 6):
- dispatcher system – DS.
- central control system – CCS
- area control system – ACS
- radio communication system – RCS
- vehicle control system – VCS

The main task of the dispatcher system is the facilitation of the monitoring of the whole system by the maintenance personnel. The PRT system is fully automatic and during regular operation no human intervention is required. In case of emergency the personnel can switch to manual control. Dispatchers monitor the current state of the traffic, power system, data transmission and proper operability of computer systems. Dispatchers can mark any component as faulty and plan maintenance tasks. The steering and controlling system is equipped with surveillance video. The video cameras are mounted on the stations and PRT vehicles. Both the stations and vehicles have systems enabling instant contact with the dispatchers In case of emergency dispatchers have direct live feedback from each vehicle. The video and voice transmissions are recorded and archived.

At the infrastructure level the system is redundant, therefore provides the continuous operability in case of failure of single components of the system. The transmission system

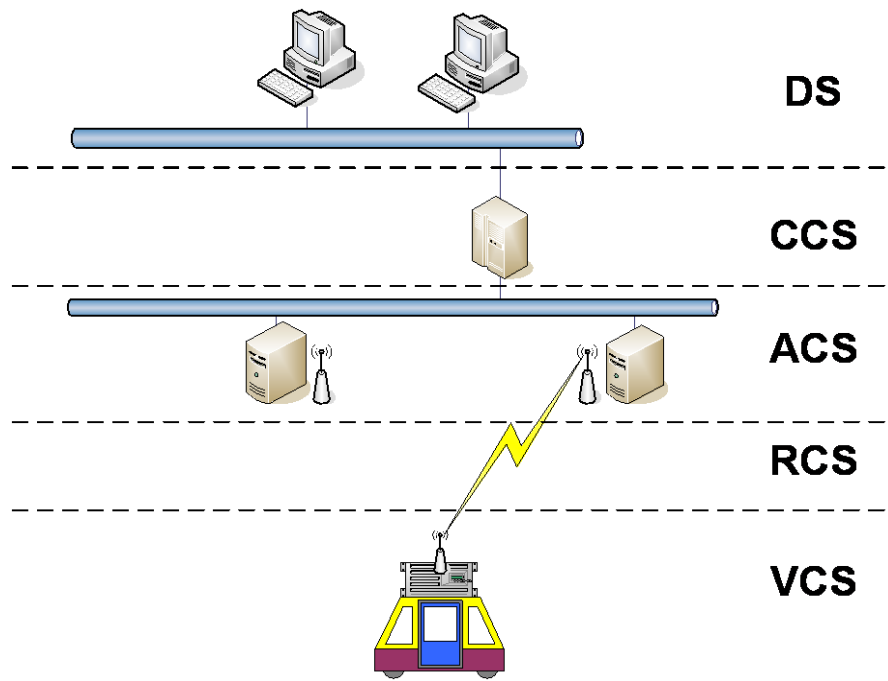

Figure 6 - Structure of the steering system of PRT

also provided redundant connection in case of failure.

Central control system is of distributed type - different functions are deployed on different steering layers. Central CCS system manages the vehicles. It sets the routes, assigns vehicles to the orders and manages empty vehicles. Performance and optimization of that part of the system is analyzed in the simulation environments. Optimization aims at the defining most efficient ways of movement in PRT network in terms of travel and waiting time as well as the energy efficiency. System react to emergency and faults situation by dynamically reassigning routes. Each change in the traffic situation is archived with a frequency of 100ms. The archived materials are available for analysis, elimination of bottleneck and errors in the system.

Area control systems have the tasks of controlling single moving vehicles, controlling the traffic on the stations and assigning right-of-way on crossroads. The whole system is divided into areas that are managed by dedicated computers. The central system collects information about location of each vehicle and distributes it to all area computers. The right-of-way algorithms are parameterized and can be adjusted according to the traffic state in specific areas. ACS does not assign or manipulate routes. In case of emergency of CCS, the ACS directs the vehicles to the nearest station or evacuation points according to predefined scenario. If the connection with specific vehicle is lost, the ACS marks corresponding area as unavailable in order to maximize the safety of all participants.

At the infrastructure level ACS is redundant. Additionally the information about movements and state of a vehicle is distributed to corresponding area computer as well as to neighbouring ones. Such setup assures that in case of emergency one of the neighbouring computers can take over the control of the area.

The main tasks of vehicle control system is communication with area control system, marking the position of the vehicle, assuring the safe distance to closest vehicle , maintaining a safe velocity, monitoring of the door controlling system, air conditioning, lightning, etc. VCS facilitates the voice and video communication between vehicles and dispatchers. Each vehicle is equipped with displays showing the current state and location of vehicle. Passenger can alter the target station at any moment and a new route will be assigned. During the trip the main display can present commercials and news information. VCS monitors the operation of the engine, breaking and power systems.

Communication between PRT vehicles and ACS is facilitated by digital transmission system. The vehicle is in range of two independent base radio stations in any point of the network. The transmission system is implemented in accordance with European standard EN-50159. Data transmission is encoded and encrypted to disable access to the systems by undesirable individuals.  Implementation of the European standards EN-50126, EN-50128, EN-50129 ensure highest safety, reliability and maintainability of the whole system. All the critical technical components are configured in  fail-safe technique, using 2 out of 2 model

## CONCLUSIONS AND FURTHER WORK

The results from simulation experiments show that the simulation environment is useful for observing and identification of effects occurring in PRT system. Also, it helps to compare various conditions of network operation, especially to find optimal conditions or to avoid traffic jams.

In the future, the comparison of various network structures, management algorithms and other features may be performed from various points of view, for example from passenger, network administrator or network maintenance engineer point of view.

The very wide set of output parameters measured give the possibility to observe the behaviour of a PRT network from that outlined or other points of view. A detailed log of events may be also obtained, which allows to build simulation statistics viewed from other points of view.

Using the simulators, various design aspects may be viewed as complex optimization tasks. For this purpose, "unit costs" of all PRT elements (length of the track, number of vehicles, number of berths in a station, number of intersections of every type etc.) and PRT services (time and distance of a travel, time of vehicle technical maintenance, cost of amortisement, etc.) must be specified. Then, optimal usage parameters (passenger waiting time, effective travel velocity, delays in comparison to optimal conditions) or maintenance parameters (daily distance travelled, full/empty travel ratio etc.) may be identified in simulation experiments.

Also, management algorithms build on other principles will be applied, i.e. decentralized

## ACKNOWLEDGMENT

The research work is carried out within the Eco‑Mobility Project that is being co-financed by the European Regional Development Fund – Operational Program for Innovative Economy. UND-POIG.01.03.01-14-154 – Project Co-ordinator Prof. W. Choromański.